# A novel strategy to control defects and secondary phases of CZTS by surfactant Potassium

Yiou Zhang, Kinfai Tse, Xudong Xiao, and Junyi Zhu

*Department of physics, the Chinese University of Hong Kong, Hong Kong SAR, China*

$Cu_2ZnSnS_4$ (CZTS) is a promising photovoltaic absorber material with earth abundant and nontoxic elements. However, the detrimental native defects and secondary phases of CSTS will largely reduce the energy conversion efficiencies. To understand the origin of these problems during the growth of CZTS, we investigated the kinetic processes on CZTS ($\bar{1}\bar{1}\bar{2}$) surface, using first principles calculations. A surface Zn atom was found to occupy the Cu site near surface easily due to a low reaction barrier, which may lead to a high $Zn_{Cu}$ concentration and a Zn-rich secondary phase. These n-type defects may create deep electron traps near the interface and become detrimental to device performance. To reduce the population of $Zn_{Cu}$ and the Zn-rich secondary phase, we propose to use K as a surfactant to alter surface kinetic processes. Improvements on crystal quality and device performance based on this surfactant are consistent with early experimental observations.

## I. INTRODUCTION

CZTS, as a promising thin-film solar-cell material, has drawn great attention in past few years [1--9]. With a zinc-blende derived kesterite structure, CZTS has a high absorption coefficient and a direct band-gap of 1.5eV, close to the optimal value for single-junction solar-cell materials [10]. Moreover, all the elements in CZTS are non-toxic and earth-abundant [11--14]. And the efficiency of a solar-cell based on an alloy of CZTS and $Cu_2ZnSnSe_4$ (CZTSe) has reached 12.6% [15]. However, the narrow stable chemical potential range of CZTS [12,13,16] leads to detrimental defects and phase inhomogeneity, which have been observed experimentally [3,10,17--19]. Also, the presence of detrimental defects in CZTS may form recombination centers that limit the energy conversion efficiency.

For thin-film solar-cell materials, modifications of surface properties are important for both material growth and photovoltaic performance [3,8,17,20--22]. However, the surface properties of CZTS are difficult to study due to the polycrystalline nature of this thin film. X-ray diffraction patterns indicate that the most favorable orientation of CZTS growth is (112)/($\bar{1}\bar{1}\bar{2}$) for both sputtering [23,24] and co-evaporation [17], and that the S-terminated ($\bar{1}\bar{1}\bar{2}$) surface has the lowest surface energy [14]. The most stable surface reconstruction under Cu poor and Zn rich condition is a (1×1) $2Zn_{Cu}$ reconstructed surface, where all the Cu sites in the first cation layer of CZTS are occupied by Zn atoms [14], resulting in the Cu-depleted region near surfaces, as observed in experiments [25].

Besides these thermodynamic properties, surface kinetics may play an important role in CZTS growth due to the low growth temperature [8,15,17], but the kinetic process that leads to the reconstruction during crystal growth remains unknown. Moreover, the effects of surface kinetic processes on defects and secondary phase formation are not studied. To propose effective surface growth techniques, it is essential to first understand the formation kinetics of the native defects and secondary phase near the surface. As direct observation of surface processes is hard in experiments, if not impossible, first principle calculations become attractive. Moreover, since first principles calculations show good consistency with experimentally observed surface properties of CZTS [14,17,23--25], they are also expected to capture the critical processes on CZTS surface.

According to our surface kinetic analysis, Zn-rich secondary phases are of special importance because it is likely to form near the CZTS ($\bar{1}\bar{1}\bar{2}$) surface kinetically, and $Zn_{Cu}$ may accumulate near the interface. To suppress the detrimental phases of $Cu_2S$, CuS, and $Cu_xSnS_y$ [26,27], the growth is usually under Cu poor and Zn rich conditions [2,11,15]. Such growth conditions also provide a thermodynamic driving force to form ZnS or other Zn-rich secondary phases. Due to the similar crystal structures and lattice constants between ZnS and CZTS, it is difficult to characterize ZnS phase in CZTS as the two X-ray diffraction patterns overlap [27]. Although a coherent interface between CZTS and ZnS may exist, and a type-I band offset between CZTS and ZnS is believed to be electronically benign [28], electronic properties of $Zn_{Cu}$ defects near the interface were not well studied and may have detrimental effects, as our further analysis shows.

To reduce the secondary phases and detrimental defects, surface modifications are appealing in the growth of semiconductors [8,14,17]. Based on understandings of surface thermodynamics and kinetics, surface growth techniques, such as surfactants, can be used to modify the surface morphology and change the surface electronic structures [23,25,26,29--31]. Surfactant elements are elements that always flow on top of the growing front [29--33]. Although surfactants are used in group III-V semiconductors [29--32,34], surfactant effect on CZTS has not been thoroughly studied. Recently it was proven experimentally that surfactant effect of Na in co-evaporation growth of CZTS can enlarge the grain size and suppress formation of ZnS secondary phase near surface [35]. This shows that surfactant effect may also play an important role in CZTS growth. Also, it was found in experiments that a small amount of K incorporation in CZTS can enhance the formation of (112) surfaces, increase the grain size, reduce ZnS secondary phase, and increase the short-circuit current [36]. Nevertheless, the physical origin remains unknown. Since it is very difficult to directly observe the surface kinetic process during the growth, first principles calculations are attractive tools to study this problem.

In this paper, to answer all the questions raised above, we studied (1) the diffusion kinetic processes of Zn, Cu near the CZTS $(\bar{1}\bar{1}\bar{2})$ surface; (2) the electronic properties of $Zn_{Cu}$ at the interface between CZTS and ZnS; and (3) the surfactant effect of K on CZTS, using first principles approaches. First, we showed that an intrinsic surface kinetic process will lead to the formation of undesirable $Zn_{Cu}$ defect and Zn-rich secondary phases. Our calculations showed that there is only a small energy barrier to form the (1×1) $2Zn_{Cu}$ reconstructed surface, whereas the recovery to stoichiometric surfaces is more difficult. These results indicate that a large number of n-type $Zn_{Cu}$ may be buried into the bulk from the surface, and Zn-rich secondary phases may be formed. Additionally, our calculations showed that despite the shallow nature of the $Zn_{Cu}$ antisite defect in bulk CZTS, such antisite defect may create deep electron traps on the interface between CZTS and ZnS. These active electron traps will greatly limit the efficiency of CZTS. Further, we showed that K can alter such intrinsic surface processes and improve the crystal quality. By changing surface electronic environment and kinetic processes, K ad-atoms can largely suppress the formation of $Zn_{Cu}$ near surface and increase the diffusion length of Zn ad-atoms. Therefore, the population of undesirable $Zn_{Cu}$ defects will be reduced and the Zn-rich secondary phases will be suppressed, consistent with experiment results [36]. Moreover, K atoms in bulk CZTS will most likely take Zn sites instead of commonly believed Cu sites [36], creating a shallow acceptor level and thus being electronically benign. Therefore, a small amount of K incorporation is in fact beneficial for the device performance of CZTS.

## II. CALCULATION DETAILS

The total energy calculations of bulks and slabs were based on Density Functional Theory [37,38] as implemented in VASP code [39,40], with a plane wave basis set [41,42]. The energy cutoff of the plane wave was set at 400eV and PBE Generalized Gradient Approximation (GGA) functional [43] was used. For surface calculations, (2×1) unit cell slabs containing 6 bi-layers of CZTS and at least 15Å vacuum are used. (3×3×1) Monkhorst-Pack [44] k-point mesh was used for integration over Brillouin zone. Pseudo-hydrogen atoms with charge $q = 1.75e$, $q = 1.5e$, and $q = e$ were added on the bottom surface to passivate dangling bonds of Cu, Zn, and Sn atoms. Atoms were relaxed until force converged to less than $0.01 eV/Å$. Transition state energy was determined by the climbing image nudged elastic band (CI-NEB) method [45]. Bulk defect properties were obtained using a supercell approach, where single defect was introduced in a 512-atom cell with Γ point only k-point sampling. Single electron energy levels of different defects and charge states were aligned using energy levels of core electrons far away from the defect. The use of large supercell allows accurate determination of both relaxation energy and defect level [46,47], matching experimental defect concentration.

## III. RESULTS AND DISCUSSIONS

### A. Surface kinetics of Zn and Cu

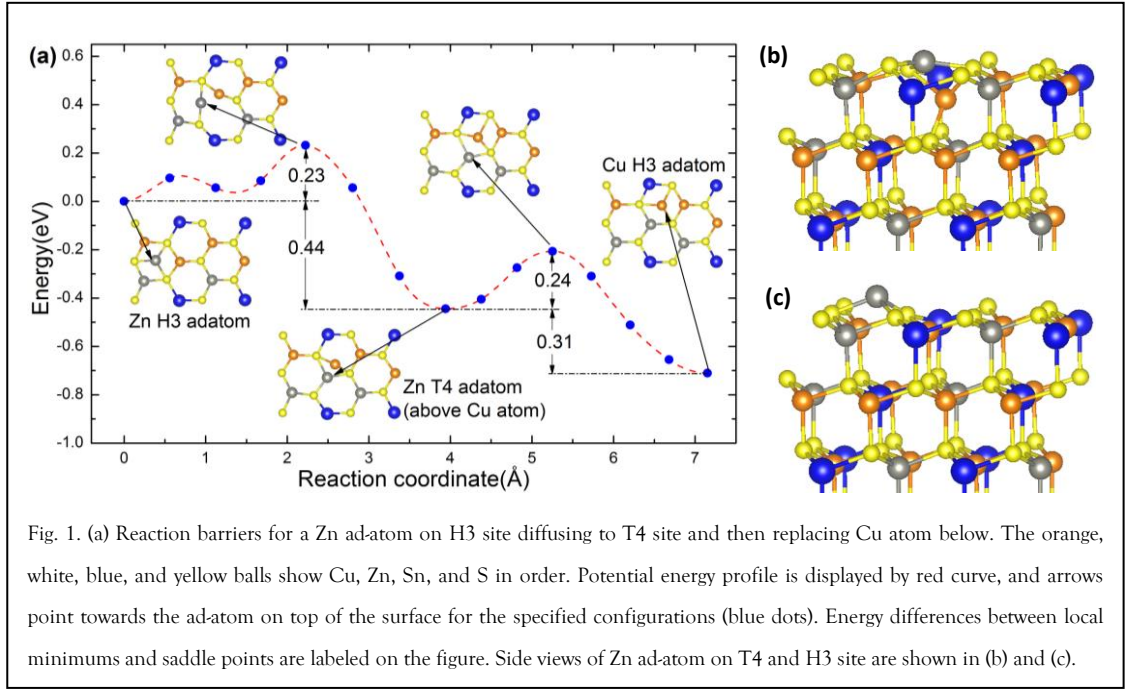

Fig. 1. (a) Reaction barriers for a Zn ad-atom on H3 site diffusing to T4 site and then replacing Cu atom below. The orange, white, blue, and yellow balls show Cu, Zn, Sn, and S in order. Potential energy profile is displayed by red curve, and arrows point towards the ad-atom on top of the surface for the specified configurations (blue dots). Energy differences between local minimums and saddle points are labeled on the figure. Side views of Zn ad-atom on T4 and H3 site are shown in (b) and (c).

We first demonstrated the kinetic process of the invasion of Cu sites in the first cation layer of the unreconstructed ($\bar{1}\bar{1}\bar{2}$) surface by ad-atom Zn. Due to low symmetry of CZTS, there are four inequivalent H3 sites and four inequivalent T4 sites, and absorption of Zn ad-atom on these sites shows different behaviors. Exhaustive testing on H3 and T4 sites shows that Zn ad-atom is stable except for T4 site above Sn atom, which will spontaneously relax towards a neighboring H3 site due to size effect of Sn atoms. Remarkably, significant distortion on Cu atom, as shown in Fig. 1(b), is observed when Zn ad-atom is on T4 site above Cu atoms. At energy minimum, Zn ad-atom is nearly co-planar with S atoms, while the Cu atom below is bonded to only 3 S atoms. Stability of such structure is due to the larger atomic size of Zn and Zn-S bonding being stronger than Cu-S bonding. Energy is gained during the stabilization of the Zn ad-atoms and the destabilization of the Cu atoms below. Comparatively, Zn ad-atoms on H3 sites, lacking such kind of energy gain mechanism, are less stable by at least 0.4eV. As a result, the adsorption of Zn ad-atom at T4 site above a Cu atom is the most stable configuration.

Since the Cu atoms below Zn ad-atoms become loosely bonded, they are more likely to diffuse away,

and the Zn ad-atoms will occupy the Cu sites and form $Zn_{Cu}$. Such diffusion will result in an exchange between the Zn ad-atoms and the Cu atoms below the Zn ad-atoms, as shown in Fig. 1(a). As can be seen from the figure, the energy barrier of the exchange process is only 0.25eV, comparable to the diffusion barrier of Zn ad-atoms on the surface. Moreover, there will be a significant energy gain when a Zn ad-atom occupied the Cu site (0.7eV compared with a Zn ad-atom on a H3 site). Therefore, Cu atoms in the first cation layer act as traps of Zn ad-atoms on the surface, and it is likely that a high population of $Zn_{Cu}$ may accumulate near the surface. Moreover, as the diffusion length of Zn ad-atoms is significantly limited by these Cu atoms, the distribution of Zn on the surface might become highly inhomogeneous. As a consequence, Zn-rich secondary phases instead of normal CZTS would be formed in the Zn-rich region, while CZTS may only be formed in some Zn-poor region. *Our calculation indicates that at the very early stage of the growth, in addition to the thermodynamic driving force, there exists a kinetic driving force to form $Zn_{Cu}$ near the surface, which may lead to formation of Zn-rich secondary phases.*

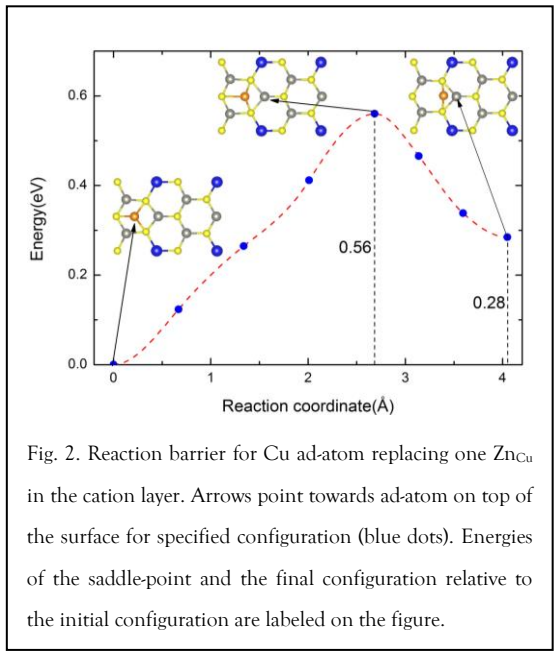

Fig. 2. Reaction barrier for Cu ad-atom replacing one $Zn_{Cu}$ in the cation layer. Arrows point towards ad-atom on top of the surface for specified configuration (blue dots). Energies of the saddle-point and the final configuration relative to the initial configuration are labeled on the figure.

Further, we studied the kinetics of Cu ad-atoms on the (1×1) $2Zn_{Cu}$ reconstructed surface. The most stable adsorption site for a Cu ad-atom is one of the H3 sites, which has the largest distance away from the Sn atoms in the cation layer. When Cu ad-atoms occupy the Cu sites, the replaced Zn atoms still favor T4 sites above the Cu atoms. As shown in Fig. 2, such a replacement is unfavorable both kinetically and thermodynamically, as the reaction barrier is more than 0.5eV and the energy of the final stage is nearly 0.3eV higher than that of the initial stage. Although the kinetic barrier may not be high under growth temperature of CZTS, the replacement is still hindered by the energy difference between the initial and final configurations. This result indicates that the (1×1) $2Zn_{Cu}$ reconstructed surface remains stable during further growth, and it is hard to remove the $Zn_{Cu}$ near the surface from both thermodynamic and kinetic aspects. Since diffusion barrier of $Zn_{Cu}$ is approximately 2eV in bulk CZTS according to our unpublished results, it is even harder to remove such antisites when they are buried in bulk. Therefore, there might be high population of $Zn_{Cu}$ in bulk CZTS due to surface kinetic processes.

### B. Electronic properties of $Zn_{Cu}$ at the interface between ZnS and CZTS

Due to the limited diffusion length of Zn ad-atom on CZTS surface and the Zn-rich condition during the growth, Zn-rich secondary phases are likely to form. Although the exact structures and composition of these secondary phases could be complex, these secondary phases can be approximately treated as doped ZnS. The ZnS secondary phase was believed to be electronically benign due to its lattice-matched structure and type-I band alignment [26]. However, such kind of benign properties exist only when the interface between CZTS and ZnS is defect-free. If this doped ZnS is formed on top of the (1×1) $2Zn_{Cu}$ reconstructed CZTS surface (which is likely during the CZTS growth), one layer of ordered defects may be formed on the interface. Such defects will be almost frozen on the interface due to the high kinetic barrier of $Zn_{Cu}$ in bulk CZTS, resulting in high defect concentrations near the interface.

To understand the electronic properties of the defects on the interface, we constructed a (1×1) unit cell along [112] direction of CZTS, containing 9 bilayers of CZTS and 9 bilayers of ZnS, to simulate this interfaces between CZTS and Zn-rich secondary phases. Although there are two in-equivalent interfaces in the super-cell, they are charge-neutral and can be regarded as fully separated. The average interface energy is only $1.5 meV/Å^2$, suggesting that the interface is almost stress-free without any charge transfer. Then $Zn_{Cu}$ on one interface were simulated, and the obtained electronic structure is shown in Fig. 3(a). As can be seen from the density-of-states plot, a deep gap state exists when one layer of defects is created on the interface. Although the defects are n-type in nature, the gap state is actually close to the VBM of CZTS, partly due to an underestimation of band-gap by GGA functional (the band-gap of CZTS is around 0.4eV from our calculations, much smaller than the band-gap of 1.5eV from experiments). Interestingly, Zn atoms have little contribution to the gap state, and the partial charge density shown in Fig. 3(c) indicates that the gap state is an anti-bonding state between Sn s-orbital and S p-orbital. These findings suggest that the defect layer can be regarded as a ZnS layer with ordered $Sn_{Zn}$ defect.

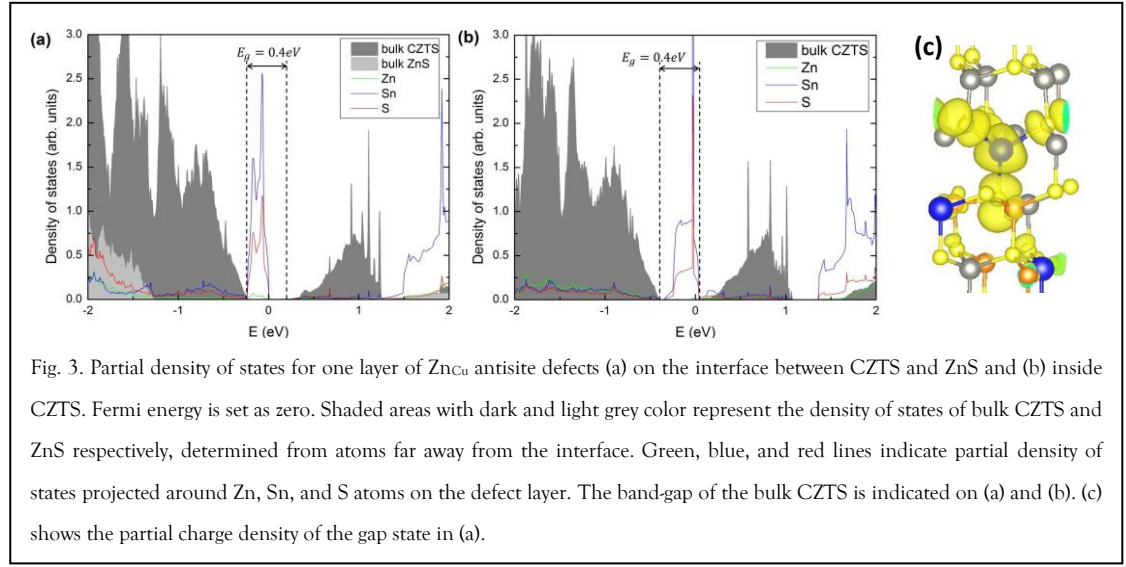

Fig. 3. Partial density of states for one layer of $Zn_{Cu}$ antisite defects (a) on the interface between CZTS and ZnS and (b) inside CZTS. Fermi energy is set as zero. Shaded areas with dark and light grey color represent the density of states of bulk CZTS and ZnS respectively, determined from atoms far away from the interface. Green, blue, and red lines indicate partial density of states projected around Zn, Sn, and S atoms on the defect layer. The band-gap of the bulk CZTS is indicated on (a) and (b). (c) shows the partial charge density of the gap state in (a).

We also studied the defect layer formed inside CZTS, as CZTS may also grow on the reconstructed surface. The calculation was done using a similar supercell with all ZnS bilayers replaced by CZTS bilayers, and the electronic structure is shown in Fig. 3(b). Similar to that on CZTS-ZnS interface, the defect layer inside CZTS also creates a gap state originating from Sn and S atoms on the defect layer. Again, the defect layer can be regarded as one layer of Sn-doped ZnS inside CZTS. Combining the calculations of $Zn_{Cu}$ defect layer on CZTS-ZnS interface and in bulk CZTS, we may conclude that under high concentration of $Zn_{Cu}$ defects, the donor electrons would transfer to neighboring Sn atom, and local structure tends to transform into Sn-doped ZnS. Effectively, some $Sn_{Zn}$ defects would be created, and the donor level becomes much deeper. Although $Zn_{Cu}$ has a relatively high defect formation energy and only creates shallow donor level in bulk CZTS [13,16], local concentration of $Zn_{Cu}$ could be high due to the surface kinetic process. As a consequence, some deep-level defects might be formed, particularly near the interface between CZTS and Zn-rich secondary phases. Eventually, they change the benign interface into a detrimental one. Although the concentration of $Zn_{Cu}$ defects may not reach such a high level, we believe that the deep nature of such defects near the interface is still qualitatively correct.

From the above calculations, we found that the charge compensating effects can significantly lower the formation energy of the n-type $Zn_{Cu}$ on the electron-poor S-terminated surface, while it is about 2eV [11--13] in the bulk. Moreover, the difference in the bond strength and atomic size between Cu and Zn results in a low reaction barrier for the formation of $Zn_{Cu}$, but a much higher barrier for the removal of such defects. This intrinsic surface processes would lead to the accumulation of $Zn_{Cu}$ near the surface, and some of them may eventually be buried into the bulk. High concentrations of such defects near interfaces with Zn-rich secondary phases or grain boundaries could create deep electron traps, significantly lowering the device performance. To suppress the formation of these undesirable defects, it is important to alter both the electronic environment and the kinetic process on CZTS surfaces.

## C. Potassium as a surfactant to reduce $Zn_{Cu}$ and ZnS

One of the effective approaches to modify the electronic environment and the kinetic process near surfaces is to incorporate surfactants. The surfactant elements are often large atoms that are difficult to diffuse into the bulk. At the same time, surfactants should help satisfy electron-counting-rule (ECR) [48--50] to lower the total energy. K atoms may serve ideally for these two purposes.

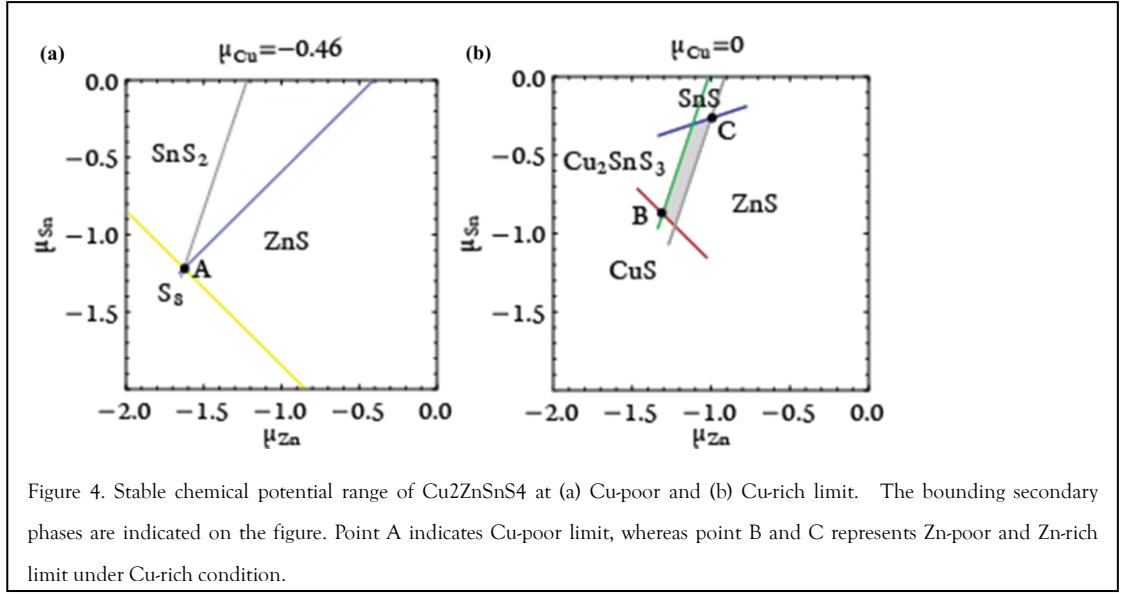

Figure 4. Stable chemical potential range of Cu2ZnSnS4 at (a) Cu-poor and (b) Cu-rich limit. The bounding secondary phases are indicated on the figure. Point A indicates Cu-poor limit, whereas point B and C represents Zn-poor and Zn-rich limit under Cu-rich condition.

Table I. Listing of formation energy (in units of eV) of K related defects at the representative points labeled in Fig. 4, with corresponding chemical potential ($\mu_{Cu}/\mu_{Zn}/\mu_{Sn}/\mu_K$) listed in parenthesis in units of eV.

|  | $K_{Cu}$ | $K_{Zn}$ | $K_{Sn}$ | $K_i$ |
|---|---|---|---|---|
| A(-0.46/-1.63/-1.22/-2.16) | 1.57 | 1.35 | 2.42 | - |
| B(0.00/-1.31/-0.88/-1.47) | 1.37 | 1.08 | 2.09 | - |
| C(0.00/-1.00/-0.27/-1.27) | 1.17 | 1.19 | 2.51 | - |

First, we calculated the formation energy of K related defects in bulk. To understand the defect properties of K substitutional and interstitial defects, we surveyed the formation energy and transition energy level of defects at all inequivalent configurations, with chemical potential range shown in Fig. 4.

Our computed stable chemical potential range and properties for selected intrinsic defects show qualitative agreement with previous work [13]. Various phases from $K_2S$ to $K_2S_6$ have been considered to determine the rich limit of K under different chemical potential conditions of CZTS. Table I listed the formation energy at selected point in the stable chemical potential range, and chemical potential of K is taken at rich limit. Even under such K-rich condition, high formation energies for all K related defects have been observed, particularly under Cu-poor condition. In contrary to Na-doped CIGS, where $Na_{Cu}$ is thermodynamically favored [51], we found that $K_{Zn}$ substitutional defect is at least as stable on the Zn-richest limit, and generally more stable over the whole chemical potential range. It is also worth noting that $K_i$ will relax simultaneously into $K_{Cu}$ and $Cu_i$, which may significantly reduce local stress, thus such configuration is unstable. The $K_{Zn}$ defect has a (-1/0) transition level of 0.045eV, possibly providing p-type carrier at room temperature. It can therefore be concluded that a high concentration of K defect in bulk CZTS is unlikely, and that trace amount of K inevitably incorporated is driven thermodynamically and kinetically into benign $K_{Zn}$ substitutional defects.

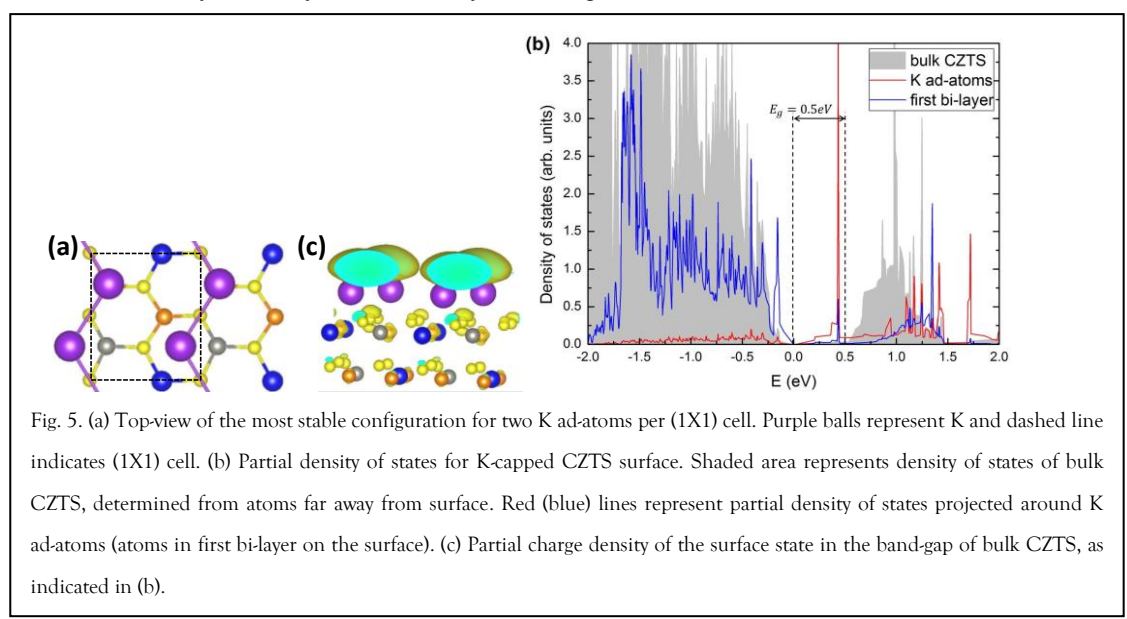

Fig. 5. (a) Top-view of the most stable configuration for two K ad-atoms per (1X1) cell. Purple balls represent K and dashed line indicates (1X1) cell. (b) Partial density of states for K-capped CZTS surface. Shaded area represents density of states of bulk CZTS, determined from atoms far away from surface. Red (blue) lines represent partial density of states projected around K ad-atoms (atoms in first bi-layer on the surface). (c) Partial charge density of the surface state in the band-gap of bulk CZTS, as indicated in (b).

Then we studied the surfactant effects of K on surfaces. Since K atoms will donate electrons to the electron-poor surface, the formation of $Zn_{Cu}$ may be suppressed thermodynamically. To verify this assumption, we studied the effects of K surfactant on the $(\bar{1}\bar{1}\bar{2})$ surface of CZTS. The two K ad-atoms were put on each (1×1) unit cell to satisfy ECR [48--50]. All possible configurations for K ad-atoms were searched, and the most stable configuration is shown in Fig. 5(a). The energy costs for exchanging K ad-atoms with Cu, Zn, and Sn atom are 0.80eV, 0.36eV, and 1.26eV, respectively, suggesting that K tends to segregate on the surface rather than incorporating into bulk. The relatively small exchange energy of K and Zn is consistent with our bulk calculations.

Since the distance between the nearest K atoms is 4.58Å, the same as that in bulk K, metallic bond may form along the 1D-chain, as confirmed by the electronic structure calculation shown in Fig. 5(b) and Fig. 5(c). Electrons in this metallic surface state are transferred to dangling bonds of S atoms, leaving an empty state above the VBM. Since all dangling bonds of S atoms are fully occupied, cation adsorption on the surface is much weaker. The adsorption energy of Zn ad-atom is 1.55eV higher than that on an un-reconstructed surface, and the Zn ad-atom is only bonded to one S atom. Also, the formation energy of $Zn_{Cu}$ is increased by 0.79eV, as the surface is no longer electron-poor. Similarly, formation energy of $Sn_{Zn}$ antisite, which is a detrimental n-type defect, is increased by 1.61eV.

Formation energies of other n-type defects are also expected to be higher on the K-capped surface. Moreover, there are no noticeable relaxations for cation atoms when Zn ad-atom is adsorbed, indicating that all atoms in CZTS bilayer remain tightly bonded. Indeed, our surface kinetics calculations, as shown in Fig. 6, revealed that diffusion barriers of Zn ad-atom on the K-capped surface are less than 0.35eV, while reaction barrier of Zn ad-atom incorporating into Cu site in the cation layer is 1.31eV, much higher than that on bare surface. Moreover, such incorporation process is thermodynamically less favorable, as the final configuration (Cu ad-atom) is 0.15eV higher than the initial one (Zn ad-atom).

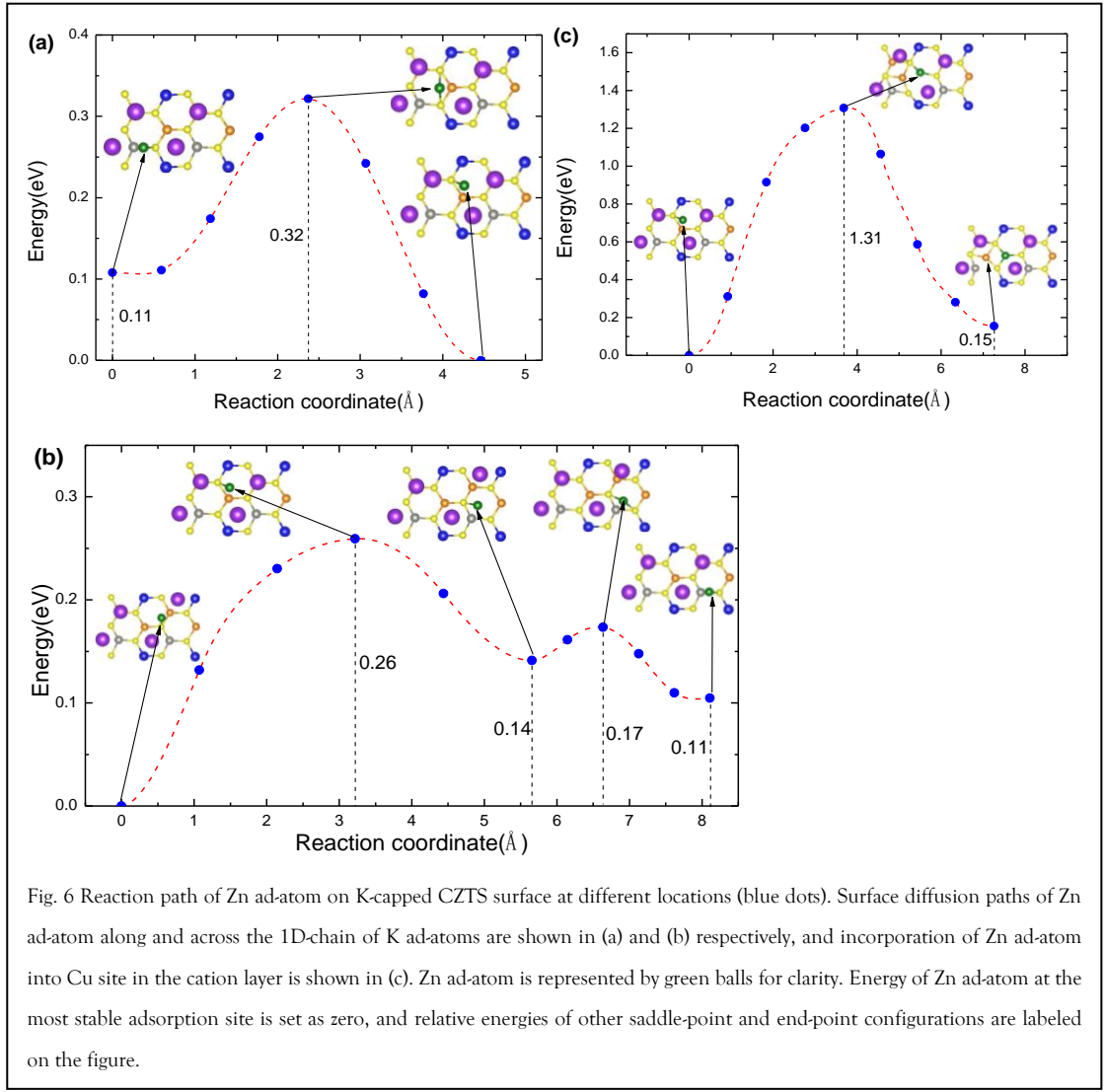

Fig. 6 Reaction path of Zn ad-atom on K-capped CZTS surface at different locations (blue dots). Surface diffusion paths of Zn ad-atom along and across the 1D-chain of K ad-atoms are shown in (a) and (b) respectively, and incorporation of Zn ad-atom into Cu site in the cation layer is shown in (c). Zn ad-atom is represented by green balls for clarity. Energy of Zn ad-atom at the most stable adsorption site is set as zero, and relative energies of other saddle-point and end-point configurations are labeled on the figure.

As the formation of $Zn_{Cu}$ near surface is suppressed both thermodynamically and kinetically, less electron traps will be formed inside CZTS and on the grain boundary, which might be the reason behind the enhancement of short-circuit current by K-doping [36]. Also, the mean diffusion length for Zn ad-atom is expected to be much longer on K-capped surfaces than that on bare surfaces. As a consequence, Zn-rich secondary phases are unlikely to form, and the grain size of CZTS could become larger, as observed in experiments [36].

## IV. SUMMARY

In summary, we have studied the intrinsic surface kinetic processes on S-terminated $(\bar{1}\bar{1}\bar{2})$ surface of CZTS, and proposed a novel surfactant effect of K on this surface. Low reaction barrier of $Zn_{Cu}$ formation on CZTS surface may result in high concentration of such defects in the bulk and the formation of Zn-rich secondary phases. These defects may create deep electron traps inside CZTS and on the interface between CZTS and ZnS, largely reducing the efficiency of CZTS solar-cell. The K atoms on CZTS surface can act as a surfactant to greatly reduce the formation of the $Zn_{Cu}$, and alter the intrinsic surface kinetic processes. As a result, the formation of $Zn_{Cu}$ and Zn-rich secondary phases can be largely suppressed, consistent with experimental results. It should be noted that the polycrystalline nature and low-symmetry of CZTS makes it difficult, if not impossible, to include all possible processes into consideration. Some important processes might be missed in our studies. Nevertheless, the surface processes we have found are likely to be critical for the quality of CZTS thin film, and surfactant effect of K is worth further investigation. Our findings also suggest that large metallic group I elements may act as ideal surfactants on electron poor surfaces to reduce the formation of detrimental native defects and secondary phases, and greatly enhance the performance of quaternary compound photo voltaic devices.

## ACKNOWLEDGEMENT

We would like to thank Bei Deng and Michael Scarpulla for helpful discussions. Computing resources were provided by the High Performance Cluster Computing Centre, Hong Kong Baptist University. This work was supported by the start-up funding at CUHK. Financial support of General Research Fund (2130490) and Research Incentive Scheme (4441641) from the Research Grants Council in Hong Kong is gratefully acknowledged.